\documentclass[10pt]{article} 

\usepackage{geometry} 

\usepackage{graphicx} 

\usepackage{float} 

\usepackage[utf8]{inputenc}
\usepackage[T1]{fontenc}
\usepackage[spanish,english]{babel}
\usepackage{lineno} 

\usepackage{authblk}
\usepackage{ulem}

\usepackage{mathrsfs}

\usepackage{amsmath}
\usepackage{amsfonts}
\usepackage{amssymb}
\usepackage[dvipsnames]{xcolor}
\usepackage{textcomp}
\usepackage{slashed}
\usepackage{hyperref}
\hypersetup{colorlinks=true}

\newcommand{\Bsphi}{\ensuremath{B_s^0 \to J/\psi \phi }}
\newcommand{\mmkk}{\ensuremath{\mu^+\mu^- K^{+}K^{-} }}
\newcommand{\Bs}{\ensuremath{B_s^0}}
\newcommand{\Bsbar}{\ensuremath{\bar{B}_{s}^{0}}}
\newcommand{\Bsh}{\ensuremath{B_s^H}}
\newcommand{\Bsl}{\ensuremath{B_s^L}}

\newcommand{\phim}{\ensuremath{\phi}}
\newcommand{\kkm}{\ensuremath{K^{+}K^{-}}}

\newcommand{\Jm}{\ensuremath{J/\psi}}
\newcommand{\Jmumu}{\ensuremath{J/\psi \to \mu^+\mu^-}}

\newcommand{\phis}{\ensuremath{\phi_s}}
\newcommand{\DGs}{\ensuremath{\Delta\Gamma_{s}}}
\newcommand{\Gs}{\ensuremath{\Gamma_{s}}}

\usepackage{fancyhdr}
\fancypagestyle{plain}{
\fancyhead[L]{Conference on Flavour Physics and CP violation (FPCP) 2020,\\ A Toxa, Spain, 8-12 June 2020}
\fancyhead[C]{}
\fancyhead[R]{}
\fancyfoot[L]{}
\fancyfoot[C]{\thepage}
\fancyfoot[R]{}
}

\begin{document}


\title{ \bf Experimental status of CPV in $B_s$ mixing}
\author{\textbf{Jhovanny Andres Mejia Guisao} \\
On behalf of the ATLAS, CMS and LHCb Collaborations\\
Physics Institute, UNIVERSIDAD DE ANTIOQUIA, Medellin, Colombia.\\
E-mail: \texttt{jhovanny.andres.mejia.guisao@cern.ch}
}

\date{}

\maketitle

\begin{center}
{ \bf Abstract} 
\end{center}

This proceeding contribution presents the latest measurement of the CP-violation and other physics parameters in the \Bsphi{} decay channel performed by the ATLAS, CMS and LHCb experiments. The measurements are based on a data set collected at the LHC in proton-proton collisions at a center-of-mass energy of 13 TeV. Besides the CP violating phase \phis{}, the width difference between the \Bs{} meson mass eigenstates \DGs{} and the average decay width \Gs{} are measured. The measured values are then combined with those collected at 8 TeV. The measurements are consistent with theoretical predictions based on the Standard Model.

\section{Introduction}

Since no direct evidence of New Physics (NP) has been found at the CERN LHC so far, precision tests of the Standard Model (SM) of particle physics have become increasingly important.
A wide program of studies of heavy flavor at the LHC, including studies of CP violation and mixing in b hadrons, is performed by the ATLAS, CMS and LHCb collaborations, and a number of recent results in this field are reviewed in this report\footnote{Copyright 2020 CERN for the benefit of the ATLAS, CMS and LHCb Collaborations. Reproduction of this article or parts of it is allowed as specified in the CC-BY-4.0 license.}. However, several other contributions in this conference include some of the new results not covered here.
In this regard, \Bs{} meson decays present copious opportunities to probe the consistency of the SM. In this proceeding, the measurement of the CP violating weak phase \phis{} and the decay width difference \DGs{} between the light (\Bsl) and heavy (\Bsh) \Bs{} meson mass eigenstates is presented. 

The \Bsphi{} channel is used to measure the CP violation phase \phis{} which is potentially sensitive to NP. In \Bsphi{} CP violation occurs due to interference between a direct decay and a decay with \Bs$-$\Bsbar{} mixing. In the SM, \phis{} is related to the elements of the Cabibbo--Kobayashi--Maskawa matrix via $\phis \simeq -2 \beta_{s}= -2arg(-V_{ts}V_{tb}^{*} / V_{cs} V_{cb}^{*})$, neglecting penguin diagram contributions. The current best determination of $-2\beta_s$ from a SM global fit to experimental data is $-36.96^{+0.72}_{-0.84}$ mrad~\cite{ref:Charles11prd}. New physics processes can introduce additional contributions to the box diagrams describing the \Bs{} mixing and can modify this \phis{} value via the contribution of beyond-the-SM particles.  So far, the NP enhancements of the mixing amplitude have been excluded by the precise measurement of the oscillation frequency. However, there is still some room on the order of statistical uncertainty. Since the numerical value of \phis{} in the SM is predicted very precisely, a deviation from the SM prediction would constitute strong evidence of physics beyond the SM.


\section{Measurement of the CP violating phase \phis{} in \Bsphi{} decays in ATLAS at 13 TeV}

The analysis presented here introduces a measurement of the \Bsphi{} decay parameters using 80.5 fb$^{-1}$ of LHC proton–proton (pp) data collected by the ATLAS detector during 2015–2017 at a centre-of-mass $\sqrt{s}$ equal to 13 TeV~\cite{ref:cpvATLAS}.

The data were collected during periods with different instantaneous luminosities, so several triggers were used in the analysis. All triggers were based on the identification of a  decay \Jmumu{}. The $K^+K^-$ pairs of the \Bsphi{} decay are selected in the vicinity of the \phim{} resonance. The four-body invariant mass spectra are fitted with signal plus background models using an unbinned extended maximum likelihood fit. 

The amplitude analyses are simultaneously performed in the helicity angles, the reconstructed mass, the measured proper decay time, the measured mass uncertainty, the measured proper decay time uncertainty, the measured transverse momentum, and the tagging probability.

To identify, or tag, the flavour of a neutral \Bs{} meson at the point of production, information is extracted using the decay of the other (or opposite) b-hadron that is produced from the pair production of $b\bar{b}$ quarks. This method is called opposite-side tagging (OST). Four types of taggers are used: two types of muons, electrons and when no lepton is found a weighted sum of the charge of the tracks in a jet associated with the opposite-side b-hadron decay is used to provide discrimination. These methods are based on the same discriminating variable called \textit{ cone charge}, defined as:

\begin{equation}
Q_{x} = \frac{\sum_{i}^{N\text{ Tracks}} q_{i} \cdot (p_{\text{T}i})^{\kappa} }{\sum_{i}^{N\text{ Tracks}} (p_{\text{T}i})^{\kappa}}\text{,}
\end{equation}

where $x=\lbrace\mu,e,jet\rbrace$ refers to muon, electron, or jet charge, respectively, and the summation is made over a selected set of tracks in a cone of size $\Delta R = \sqrt{(\Delta \phi)^2 + (\Delta \eta)^2}<0\text{.}5$ around the lepton or jet direction. The constant $\kappa$ is found empirically for each tagging method.
A probability $P(B|Q_x)$ is constructed, which is defined as the probability that a B meson is produced in a state containing a $\bar{b}-$quark, given the value of the cone charge $Q_x$. The tagging power of a particular tagging method is then defined as $T_x = \sum_{i} \epsilon_{xi} \cdot D^2(Q_{xi})$ where the sum is over the probability distribution in intervals of the cone charge variable, $\epsilon_x$ is the tag efficiency, and $D(Q_{xi})$ is the dilution. The efficiency,  $\epsilon_x$, of an individual tagging method is defined as the number of signal events tagged by that method divided by the
total number of signal events in the sample. The purity of a particular flavour tagging method, called the dilution, is defined as $D(Q_{xi}) = 2P(B|Q_x)-1$. An effective dilution, $D(Q_{xi}) = \sqrt{T_x/\epsilon_x}$ , is calculated from the measured tagging power and efficiency. A summary of the tagging performance for each method is given in Table~\ref{table:ATLAS1}.

\begin{table}[h]
    \centering
    \caption{Summary of tagging performance for the different flavour tagging methods. Taken from~\cite{ref:cpvATLAS}.}
    \begin{tabular}{lccc}
        \hline
        \hline        
        Tag method & $\epsilon_x$[\%] & $D_{x}$[\%]  & $T_x$[\%] \\        
        \hline 
        Tight muon & $4.50 \pm 0.01$ & $ 43.8 \pm 0.2$ & $ 0.862 \pm 0.009 $ \\
        Electron & $1.57 \pm 0.01$ & $ 41.8 \pm 0.2$ & $ 0.274 \pm 0.004 $ \\
        Low-$p_{T}$ muon & $3.12 \pm 0.01$ & $ 29.9 \pm 0.2$ & $ 0.278 \pm 0.006 $ \\
        Jet & $12.04 \pm 0.02$ & $ 16.6 \pm 0.1$ & $ 0.334 \pm 0.006 $ \\        
        \hline
        Total & $21.23 \pm 0.03$ & $28.7 \pm 0.1 $ & $1.75 \pm 0.01 $ \\
        \hline 
        \hline
    \end{tabular}
    \label{table:ATLAS1}
\end{table}

The results of the main physics parameters obtained from the fit are given in Table~\ref{table:ATLAS2}.

\begin{table}[H]
    \centering
    \caption{Fitted values for the physical parameters of interest with their statistical and systematic uncertainties. Taken from~\cite{ref:cpvATLAS}.}
    \begin{tabular}{cccc}
        \hline
        \hline        
        Parameter & value & Statistical   &  Systematic \\    
         &  &  uncertainty  &   uncertainty \\                    
        \hline 
        \phis{} [rad] & $-0.081$ & $ 0.041 $ & $0.020$ \\
        \DGs{} [ps$^{-1}$] & $0.0607$ & $ 0.0047 $ & $0.0022$ \\
        \Gs{} [ps$^{-1}$] & $0.6687$ & $ 0.0015 $ & $0.0018$ \\        
        \hline 
        \hline
    \end{tabular}
    \label{table:ATLAS2}
\end{table}

\begin{figure}[h]
\centering
\resizebox{0.75\textwidth}{!}{%
\includegraphics{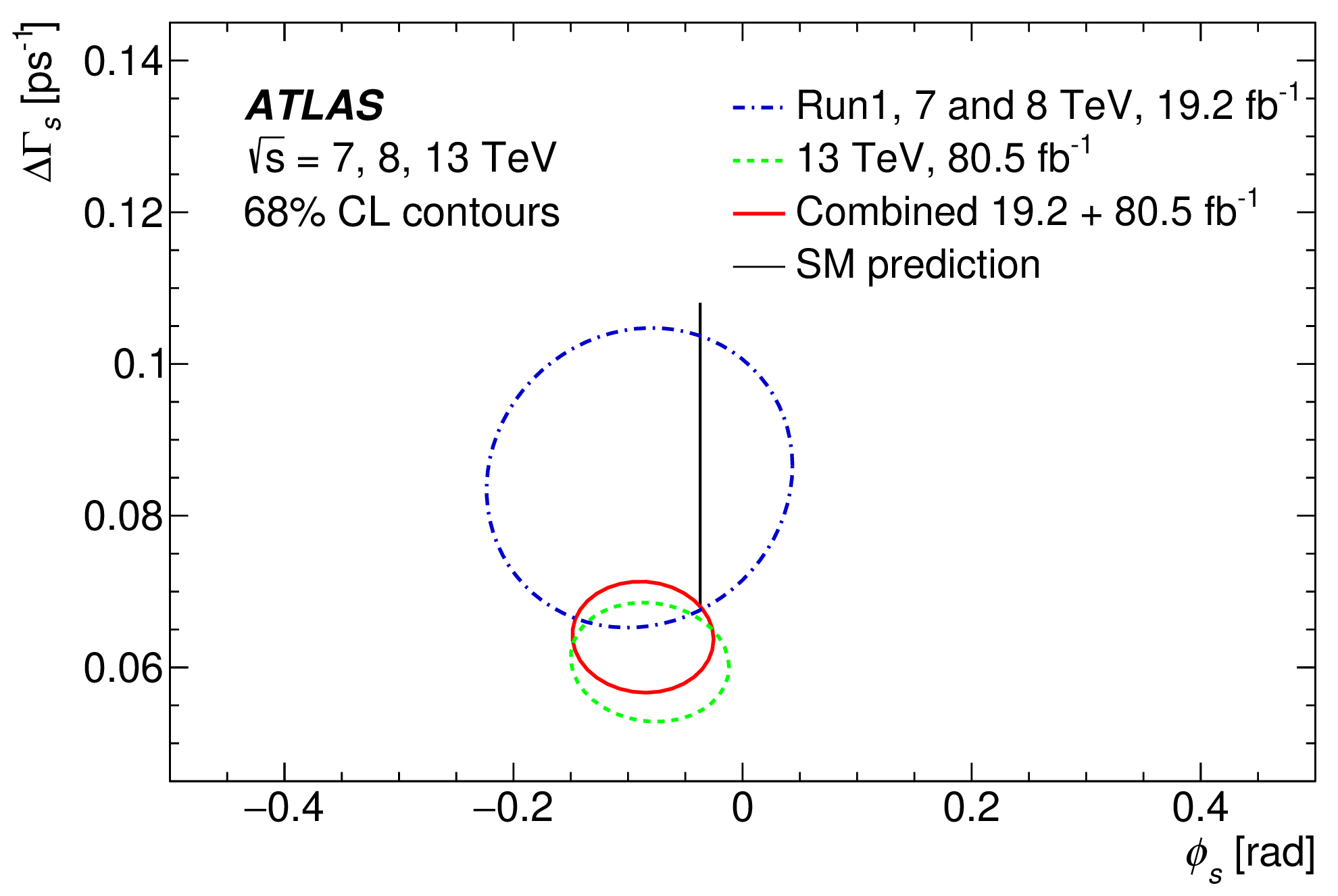}}
\caption{Contours of 68\% confidence level in the \phis-\DGs{} plane, showing ATLAS results for 7 TeV and 8 TeV data (blue dashed-dotted curve), for 13 TeV data (green dashed curve) and for 13 TeV data combined with 7 TeV and 8 TeV (red solid curve) data. The Standard Model prediction is shown as a very thin black rectangle. In all contours the statistical and systematic uncertainties are combined in quadrature and correlations are taken into account. Taken from~\cite{ref:cpvATLAS}.}
\label{fig:ATLAS1}
\end{figure}

The obtained parameters are combined with results from ATLAS 8 TeV measurements~\cite{ref:cpvATLAS_8tev} using a Best Linear Unbiased Estimator method (BLUE). This method uses the measured values and uncertainties of the parameters as well as the correlations between them. The combined values of parameters are shown in Table~\ref{table:ATLAS3}. The two-dimensional likelihood contours in the \phis-\DGs{} plane for the ATLAS result based on 7 TeV and 8 TeV data, the result from 13 TeV data, and the combined results are shown in Figure~\ref{fig:ATLAS1}.

\begin{table}[H]
    \centering
    \caption{Values of the physical parameters extracted in the combination of 13 TeV results with those obtained from 7 TeV and 8 TeV data. Taken from~\cite{ref:cpvATLAS}.}
    \begin{tabular}{cccc}
        \hline
        \hline        
        Parameter & value & Statistical   &  Systematic \\    
         &  &  uncertainty  &   uncertainty \\                    
        \hline 
        \phis{} [rad] & $-0.087$ & $ 0.036 $ & $0.019$ \\
        \DGs{} [ps$^{-1}$] & $0.0641$ & $ 0.0043 $ & $0.0024$ \\
        \Gs{} [ps$^{-1}$] & $0.6697$ & $ 0.0014 $ & $0.0015$ \\        
        \hline 
        \hline
    \end{tabular}
    \label{table:ATLAS3}
\end{table}

\section{LHCb Updated measurement of time-dependent CP violating observables in \Bsphi{} decays}

The measurement reported is based on a data sample of proton–proton collisions recorded by the LHCb experiment at a centre-of-mass energy of 13 TeV in 2015 and 2016, corresponding to an integrated luminosity of 0.3 fb$^{-1}$ and 1.6 fb$^{-1}$, respectively~\cite{ref:cpvLHCb}.

The \Bs{} candidates are reconstructed by combining the \Jm{} candidate with the \kkm{} pair. The invariant mass of the  \kkm{} pair, m(\kkm{}), must be in the range 990–1050 MeV. The selection of signal events over the combinatorial background is performed with a BDT using kinematic variables. The m(\mmkk ) distribution is fitted using an unbinned maximum-likelihood fit to the sample.

The amplitude analyses are simultaneously performed in the helicity angles and decay time requiring a careful study of angular and decay-time efficiencies, time resolution, and flavor tagging. The decay-time resolution is found to be $\sigma_{\text{eff}}$ $\simeq$ 45.54 fs and a 30\% higher tagging power than 8 TeV has been achieved. The decay-time and angular efficiencies are estimated with simulation and corrected with data methods. It is worth saying, in this analysis the quantity \Gs{}-$\Gamma_{d}$ is measured instead of \Gs{}.

The determination of the initial flavor of the \Bs{} meson, called tagging, used two classes of algorithms: the opposite side
(OS) tagger and the same-side kaon (SSK) tagger. The tagging algorithms each provide a flavor-tagging decision and an estimate of the probability that the decision is incorrect (mistag) for each reconstructed \Bs{} candidate. The mistag probability, $w$,  is defined in the range from 0 to 0.5. Each tagger has a corresponding tagging power given by  $\epsilon_{tag} D^2$, where $\epsilon_{tag}$  is the fraction of tagged candidates and $D = 1-2w$ is the dilution induced on the amplitude of the \Bs{} oscillation. The combined estimated mistag probability and the corresponding uncertainties are obtained by combining the individual calibrations for the OS and SSK tagging and propagating their uncertainties. The effective tagging power and efficiency for these both OS and SSK tagged candidates is given in Table~\ref{table:LHCb1}.

\begin{table}[h]
    \centering
    \caption{Overall tagging performance. Taken from~\cite{ref:cpvLHCb}.}
    \begin{tabular}{lccc}
        \hline
        \hline        
        Category & $\epsilon_{tag}$(\%) & $D^{2}$  & $\epsilon_{tag} D^2$(\%) \\        
        \hline 
        OS only & $11.4$ & $ 0.078$ & $ 0.88 \pm 0.04$ \\
        SSK only & $42.6$ & $ 0.032$ & $ 1.38 \pm 0.30$ \\
        OS \& SSK & $23.8$ & $ 0.104$ & $ 2.47 \pm 0.15$ \\
        \hline
        Total & $77.8$ & $0.061$ & $4.73 \pm 0.34 $ \\
        \hline 
        \hline
    \end{tabular}
    \label{table:LHCb1}
\end{table}

The measured values are:

\begin{equation}
\begin{split}
\phis = & \, -0.083 \pm 0.041 \pm 0.006 \text{ rad} \, , \\
\Gs-\Gamma_{d} = & \, -0.0041 \pm 0.0024 \pm 0.0015 \text{ ps}^{-1} \, , \\
\DGs = & \, 0.077 \pm 0.008 \pm 0.003 \text{ ps}^{-1}  \, ,
\end{split}
\end{equation}

These results are combined with related 8 TeV LHCb measurements, taking into account
all statistical correlations, all systematic errors and their correlations, and correlations between different run periods.
In the 8 TeV measurement, \Gs{} was measured instead of \Gs{}-$\Gamma_{d}$, hence a linear transformation is taken into account in the combination, constraining $\Gamma_{d}$ to the known value~\cite{ref:Gd}. The combined results are

\begin{equation}
\begin{split}
\phis = & \, -0.080 \pm 0.032 \text{ rad} \, , \\
\Gs = & \, 0.6570 \pm 0.0023 \text{ ps}^{-1} \, , \\
\DGs = & \, 0.0784 \pm 0.0062  \text{ ps}^{-1}  \, .
\end{split}
\end{equation}

\section{CMS Measurement of the CP violating phase \phis{} in the $\Bsphi \to \mu^{+}\mu^{-}K^{+}K^{-}$ channel in proton-proton collisions at $\sqrt{s}=$13 TeV}

The $CP$ violating weak phase \phis{}, and the decay width difference \DGs{} between the light and heavy \Bs{} mass eigenstates are measured with the CMS detector at the LHC in a sample of reconstructed $\Bsphi \to \mu^{+}\mu^{-}K^{+}K^{-}$ decays. The measurement is based on a data set corresponding to an integrated luminosity of $96.4\;\mathrm{fb}^{-1}$, collected in proton-proton collisions at a center-of-mass energy of $13\;\mathrm{TeV}$ in 2017-2018~\cite{ref:cpvCMS}\footnote{The preliminary results shown in this contribution are superseded in this paper arXiv:2007.02434[hep-ex], Submitted to Phys. Lett. B.}. Compared to the previous measurement~\cite{ref:cpvCMS_8tev}, this analysis benefits from a data set collected at higher center-of-mass energy, as well as from a novel opposite-side (OS) muon flavor tagger, taking advantage of machine learning techniques, which allowed reaching unprecedented tagging power.  It also uses a specialized trigger, requiring an additional muon, which improves the tagging efficiency at the cost of the reduced number of signal events.

A trigger optimized for the detection of b hadrons decaying to \Jm{} mesons is used to collect the data sample. Besides, a third muon is required. This additional muon is important to infer the flavor of the \Bs{} meson at production.
Selections have been optimized using the \textit{TMVA} package. The \phim{} candidates are selected if the track pair, \kkm{}, of the \Bsphi{} decays has an invariant mass within $10$ MeV of the world-average $\phi(1020)$ meson mass.

An unbinned extended maximum-likelihood  is performed by including the information on the \Bs{} candidate invariant mass, the three decay angles of the reconstructed \Bs{} candidate, the flavor tag decision, the mistag fraction, the proper decay length of the \Bs{} candidate, and its proper decay length uncertainty.

The flavor of the \Bs{} candidate at production is determined with an OS flavor tagging algorithm. The tagging efficiency $\epsilon_{tag}$ is defined as the fraction of the total number of events which are tagged. The mistag fraction $w_{tag}$ is defined as the ratio between the number of wrongly tagged events to the total number of tagged events, and it is used to compute the dilution $D = 1-2w_{tag}$, which is a measure of the performance degradation due to mistagged events.  The tagging power $P_{tag} = \epsilon_{tag}D^{2}$ is the effective tagging efficiency. The final flavor tagging performance is shown in Table~\ref{table:CMS1}.

\begin{table}[h]
    \centering
    \caption{OS muon tagger performance using events in 2017 and 2018 data sets. Taken from~\cite{ref:cpvCMS}.}
    \begin{tabular}{lccc}
        \hline
        \hline        
        Data set & $\epsilon_{tag}$(\%) & $w_{tag}$(\%)  & $P_{tag}$(\%) \\        
        \hline 
        2017 & $45.7 \pm 0.1$ & $27.1 \pm 0.1$ & $9.6 \pm 0.1$ \\
        2018 & $50.9 \pm 0.1$ & $27.3 \pm 0.1$ & $10.5 \pm 0.1$ \\        
        \hline 
        \hline
    \end{tabular}
    \label{table:CMS1}
\end{table}

The results of the fit with their statistical and systematic uncertainties are given in Table~\ref{table:CMS2}.

\begin{table}[H]
    \centering
    \caption{Results of the fit to data with the statistical and systematic uncertainties. Taken from~\cite{ref:cpvCMS}.}
    \begin{tabular}{cccc}
        \hline
        \hline        
        Parameter & Fit result & Stat. uncer.   &  Syst. uncer. \\    
        \hline 
        \phis{} [rad] & $-0.011$ & $ \pm0.050 $ & $\pm0.010$ \\
        \DGs{} [ps$^{-1}$] & $0.114$ & $ \pm0.014 $ & $\pm0.007$ \\
        \Gs{} [ps$^{-1}$] & $0.6531$ & $ \pm0.0042 $ & $\pm0.0024$ \\        
        \hline 
        \hline
    \end{tabular}
    \label{table:CMS2}
\end{table}

The results presented are further combined with the earlier CMS result at the center-of-mass energy of $\sqrt{s}=8$ TeV~\cite{ref:cpvCMS_8tev}, using the \textit{IMINUIT} python interface of the \textit{MINUIT2} minimization package. The combined results for the CP violating phase and the lifetime difference between the two mass eigenstates are:

\begin{equation}
\begin{split}
\phis = & \, -0.021 \pm 0.045 \text{ rad} \, , \\
\DGs = & \, 0.1074 \pm 0.0097 \text{ ps}^{-1}  \, .
\end{split}
\end{equation}

\begin{figure}[H]
\centering
\resizebox{0.75\textwidth}{!}{%
\includegraphics{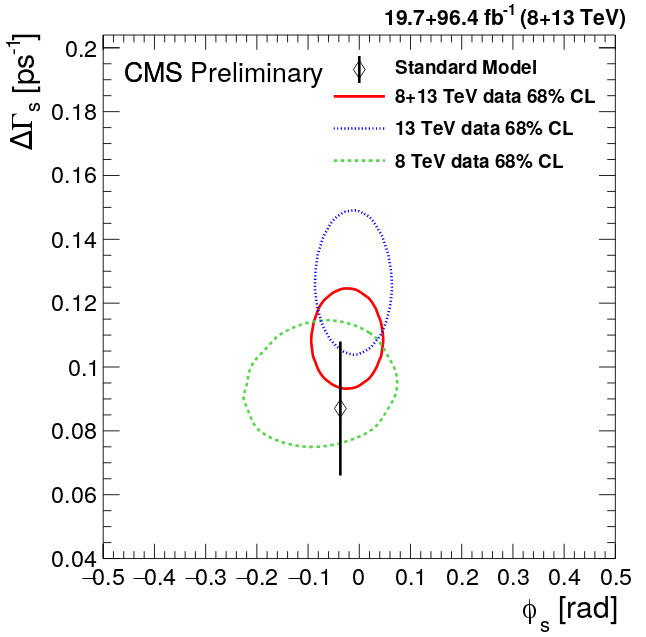}}
\caption{The two-dimensional likelihood contours at 68\% CL in the \phis-\DGs{} plane, for the CMS 8~TeV (dashed green line), 13~TeV (dashed blue line), and the combined (solid red line) results. The SM prediction is shown as the black rectangle. Taken from~\cite{ref:cpvCMS}.}
\label{fig:CMS1}
\end{figure}

The two-dimensional \phis-\DGs{}  likelihood contours at 68\% confidence level (CL) for the individual and combined results, as well as the SM prediction, are shown in Figure~\ref{fig:CMS1}. The results are in agreement with each other and with the SM predictions.

\section{Acknowledgements}

The author wants to acknowledge the support of COLCIENCIAS 
 and Universidad de Antioquia.

\end{document}